\documentclass[journal=jcisd8,manuscript=article,layout=twocolumn]{achemso}
\usepackage[T1]{fontenc} 
\usepackage{amsmath}
\usepackage[version=4]{mhchem}
\usepackage{tabularx}
\usepackage[hidelinks]{hyperref}
\usepackage{listings}
\usepackage{xcolor}
\definecolor{codegreen}{rgb}{0,0.6,0}
\definecolor{codegray}{rgb}{0.5,0.5,0.5}
\definecolor{codepurple}{rgb}{0.58,0,0.82}
\definecolor{backcolour}{rgb}{0.95,0.95,0.92}
 
\lstdefinestyle{mystyle}{
    backgroundcolor=\color{backcolour},   
    commentstyle=\color{codegreen},
    keywordstyle=\color{magenta},
    numberstyle=\tiny\color{codegray},
    stringstyle=\color{codepurple},
    basicstyle=\ttfamily\scriptsize,
    breakatwhitespace=false,         
    breaklines=true,                 
    captionpos=b,                    
    keepspaces=true,                 
    numbers=left,                    
    numbersep=5pt,                  
    showspaces=false,                
    showstringspaces=false,
    showtabs=false,                  
    tabsize=2
}
 
\usepackage[flushleft]{threeparttable}

\author{Yunqi Shao}
\affiliation[]{Department of Chemistry-\AA ngstr\"om Laboratory, Uppsala
  University, L\"agerhyddsv\"agen 1, BOX 538, 75121 Uppsala, Sweden}
\author{Matti Hellstr\"om}
\affiliation[]
{Software for Chemistry and Materials B.V., De Boelelaan 1083, 1081HV Amsterdam, The Netherlands}
\email{* hellstrom@scm.com}
\author{Pavlin D. Mitev}
\affiliation[]{Department of Chemistry-\AA ngstr\"om Laboratory, Uppsala
  University, L\"agerhyddsv\"agen 1, BOX 538, 75121 Uppsala, Sweden}
\author{Lisanne Knijff}
\affiliation[]{Department of Chemistry-\AA ngstr\"om Laboratory, Uppsala
  University, L\"agerhyddsv\"agen 1, BOX 538, 75121 Uppsala, Sweden} 
\author{Chao Zhang}
\affiliation[]{Department of Chemistry-\AA ngstr\"om Laboratory, Uppsala
  University, L\"agerhyddsv\"agen 1, BOX 538, 75121 Uppsala, Sweden}
\email{* chao.zhang@kemi.uu.se}

\title[]{PiNN: A Python Library for Building Atomic Neural Networks of Molecules and Materials}

\begin{document}

\begin{abstract}
  Atomic neural networks (ANNs) constitute a class of machine learning methods for predicting potential energy surfaces and physico-chemical properties of molecules and materials. Despite many successes, developing interpretable ANN architectures and implementing existing ones efficiently are still challenging. This calls for reliable, general-purpose and open-source codes. Here, we present a python library named PiNN as a solution toward this goal. In PiNN, we designed a new interpretable and high-performing graph convolutional neural network variant, PiNet, as well as implemented the established Behler-Parrinello high-dimensional neural network. These implementations were tested using datasets of isolated small molecules, crystalline materials, liquid water and an aqueous alkaline electrolyte. PiNN comes with a visualizer called PiNNBoard to extract chemical insight ``learned'' by ANNs, provides analytical stress tensor calculations and interfaces to both the Atomic Simulation Environment and a development version of the Amsterdam Modeling Suite. Moreover, PiNN is highly modularized which makes it useful not only as a standalone package but also as a chain of tools to develop and to implement novel ANNs. The code is distributed under a permissive BSD license and is freely accessible at \href{https://github.com/Teoroo-CMC/PiNN/}{https://github.com/Teoroo-CMC/PiNN/} with full documentation and tutorials.
\end{abstract}

\section{Introduction}
\label{sec:introduction}

One major task of computational chemistry is to map the structure of a molecule or a material to its property, i.e. $f: \{\vec{x}_i, Z_i\} \rightarrow P$. When $P$ is the total energy, then the task is to devise computational methods to find approximate solutions to the Schr\"odinger equation, as Dirac foresaw in his 1929 account~\cite{London:1929cq} and what generations of computational and theoretical chemists have been devoted to. What is even more challenging is to do the reverse $f:  P \rightarrow \{\vec{x}_i, Z_i\}$, i.e., to propose new structures which have properties of particular value. 

To address these challenges, machine learning (ML) has attracted considerable attention and efforts in computational chemistry and materials discovery~\cite{Bartok:2017hz, AspuruGuzik:2018df, Butler:2018fl}, and many different types of ML methods have been successfully applied in those areas. In this work, we will focus on atomic neural networks (ANNs), that have been very successful in predicting physico-chemical properties, approximating potential energy surfaces (PES)~\cite{behlerGeneralizedNeuralNetworkRepresentation2007, schuttQuantumChemicalInsightsDeep2017} and allowing for simulations of large-scale systems with the accuracy of reference electronic structure calculations but at only a fraction of the computational cost.

In spite of this great promise and present success, 
the development of ANNs is not straightforward. ANNs must preserve rotational, translational and permutational invariances of the system,
which was recognized in the early days of ANN development~\cite{Gassner:1998fx}. 
Besides using functions of internal coordinates, which are rotationally and translationally invariant, 
a symmetrization process was proposed to preserve also permutational invariance. 
However, such models could only be applied to systems of a given size and large-scale simulations were not possible.

Behler-Parrinello neural networks (BPNNs)~\cite{behlerGeneralizedNeuralNetworkRepresentation2007}, or high-dimensional neural network potentials, 
introduced the ansatz of partitioning the total potential energy of the system
into effective atomic contributions. Not only
does the atomic energy ansatz enable applying
a trained ANN to systems of different sizes, but
it also transforms the problem of describing the full system to that of describing the local chemical environment of each atom~\cite{behlerAtomcenteredSymmetryFunctions2011,behlerConstructingHighdimensionalNeural2015}. BPNNs have been successfully constructed for a wide range of molecules~\cite{smithANI1ExtensibleNeural2017,Gastegger-ChemSci-2017,Yao:2018fy, Schran-JCP-2018} and materials~\cite{Khaliullin-NatMat-2011,Artrith-CMS-2015,Morawietz8368,Hellstrom-ChemSci-2019}.

The BPNN architecture relies on calculating fingerprints of the atomic environment using a set of symmetry functions, that need to be selected before the fitting procedure can begin~\cite{behlerGeneralizedNeuralNetworkRepresentation2007}.
In contrast, the features are automatically learned via a feature hierarchy rather than handcrafted in convolutional neural networks, one of the most successful end-to-end techniques in the field of deep learning which won the ImageNet competition in 2012~\cite{Krizhevsky:2012wl}. Because molecules and materials can be viewed as fully connected graphs, this led to the development of graph convolution neural networks (GCNN) in atomic systems~\cite{Duvenaud:2015ww, Kearnes:2016dt,schuttQuantumChemicalInsightsDeep2017}. Hierarchical atomic features can be obtained by applying multi-stage concatenated convolution operations and this leads to impressive performance for a variety of systems ~\cite{schuttQuantumChemicalInsightsDeep2017, schuttSchNetDeepLearning2018, xieCrystalGraphConvolutional2018,lubbersHierarchicalModelingMolecular2018, Unke:2019bp, chenGraphNetworksUniversal2019,Zubatyuk:2019gp} . Among those, SchNet~\cite{schuttSchNetDeepLearning2018} is a leading example for extending GCNN methods to the modeling of both molecules and materials. 

Despite these progresses, developing interpretable ANN architectures~\cite{Chen-JCTC-2018,Schutt-preprint-2018}, and implementing existing ones efficiently are still challenging. Therefore, to promote the application of ANNs in computational chemistry and materials science communities, reliable, general-purpose and open-source codes are needed~\cite{Khorshidi:2016fua, Artrith:2016gca, WANG2018178, Singraber:2019dy, Schutt:2019fd}. Here, we present a python library named PiNN as a solution toward this goal. 

PiNN supports both BPNNs and GCNNs. The unique features of PiNN are, for example, i) that it contains a new type interpretable and high-performing GCNN variant, namely PiNet, and ii) that the interpretation of GCNN in PiNN is given by visualizing feature activation using a user-friendly JavaScript plugin PiNNBoard, instead of doing dimension reduction of embedding vectors~\cite{xieCrystalGraphConvolutional2018,chenGraphNetworksUniversal2019,Zubatyuk:2019gp} or generating 3D potential map of a test particle~\cite{schuttSchNetDeepLearning2018}. Moreover, PiNN provides analytical stress tensor calculations for lattice optimizations and constant pressure MD simulations.

In the following, we first introduce PiNN's representation and abstraction of ANNs with focus on PiNet --- an interpretable GCNN architecture we developed. Then, we discuss implementation and package features of PiNN. After that, we present different case studies for molecules, crystalline materials and liquids. Finally, we conclude with an outlook. 

\section{PiNN's representation of atomic neural networks}
\label{sec:representation}

\subsection{Representing local chemical environments}

The many-body expansion decomposes the total potential energy $E_\text{tot}$ of a system of $N$ atoms into $n$-body terms $E^{(n)}$:

\begin{equation}
\begin{split}
    E_\text{tot} &= \sum_{i}^{N} E^{(1)}(\vec{x}_i;Z_i) + \sum_{j>i}^{N} E^{(2)}(\vec{x}_i,\vec{x}_j;Z_i, Z_j) \\
    &+ \sum_{j>i}^{N}\sum_{k>j}^{N} E^{(3)}(\vec{x}_i, \vec{x}_j, \vec{x}_k; Z_i, Z_j, Z_k) + \cdots  \\
     &= \sum_i^N E_i(\vec{x}_1,\cdots, \vec{x}_N; Z_1, \cdots, Z_N)
     \label{atomE}
\end{split}
\end{equation}
where $\vec{x}_i$ is the atomic position and $Z_i$ is the atomic number. In ANNs discussed in this work, $E_\text{tot}$ by construction equals the sum of all effective atomic energies $E_i$. Note that one may also apply this construction to other atomic properties, such as partial charges.

In the construction of ANNs, descriptors which satisfy certain conditions such as symmetry invariance are needed. Two categories of descriptors have been developed to represent local chemical environments around atoms. One is inspired by the many-body expansion to include radial and angular terms~\cite{Huang:2016gh} (but this does not mean only two-body and three-body information are captured), e.g. atom-centered symmetry functions~\cite{behlerGeneralizedNeuralNetworkRepresentation2007, Gastegger:2018ec},  Faber-Christensen-Huang-Lilienfeld (FCHL) representation~\cite{Faber:2018fva} and local reference frames~\cite{Zhang:2018kza}.
 The other is using the expansion of atomic density in terms of orthogonal radial functions and spherical harmonics~\cite{Willatt:2019jc}, e.g. smooth overlap of atomic positions (SOAP)~\cite{Bartok:2013cs}. However, as pointed out recently~\cite{Drautz:2019fwa,Willatt:2019jc}, these phenomenological categories are indeed connected and one can systematically build up many-body representations from atomic cluster expansion in terms of single-bond basis functions~\cite{Drautz:2019fwa}, tensor product of symmetrized two-body dyad~\cite{Willatt:2019jc} or finite powers of the two-body base kernel~\cite{Glielmo:2018bm}.
 
 The key insight of these recent works is that the full set of bond vectors originating from a given center atom in an atomic configuration of molecules or materials contains the complete information about the corresponding structure and a many-body representation can be generated through interactions of bond vectors. In fact, the same inspiration is shared in GCNNs, where the starting point is a directed graph of the structure with annotated bonds as edges and the many-body representation in the latent space is generated from a series of graph convolution blocks (See Section Pairwise Interactions and Interaction Pooling for more discussions).

\begin{figure*}
  \centering
  \includegraphics[width=0.8\textwidth]{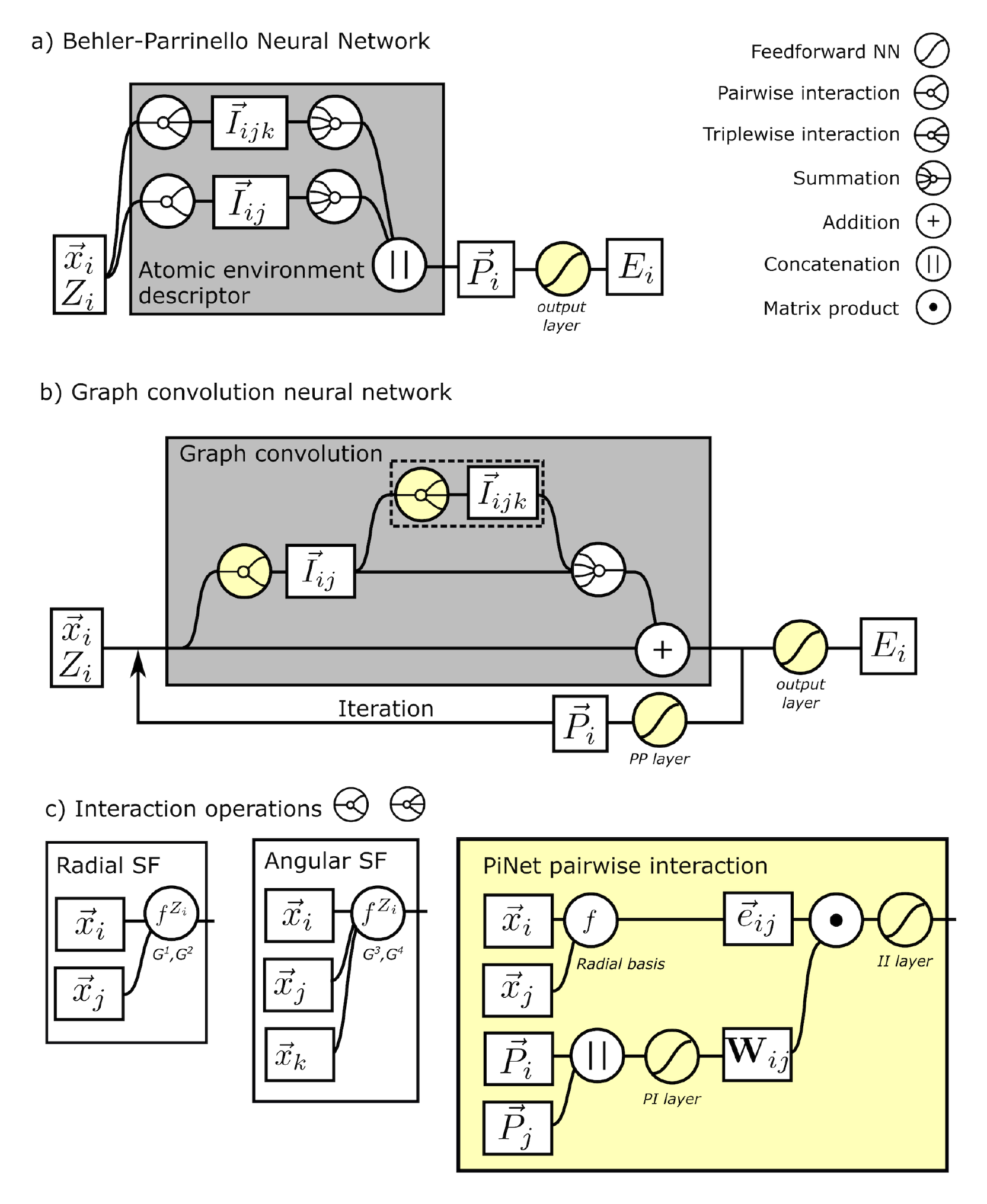}
  \caption[networks]{Illustration of the abstractions of BPNN and GCNN  frameworks in PiNN, for calculating atomic energies $E_i$ (Eq.~\ref{atomE}). The operations containing
    trainable variable are filled with yellow color. Operations inside the dashed box are not yet implemented in PiNN but extendable. See Section ``Pairwise Interaction and Interaction Pooling'' for more explanations.}
  \label{fig:networks}
\end{figure*}

\subsection{Behler-Parrinello and graph convolutional neural networks}

Here, we will briefly describe two important ANN architectures relevant to PiNN.

In BPNNs, the chemical environment around an atom is represented by a vector of symmetry functions~\cite{behlerGeneralizedNeuralNetworkRepresentation2007, behlerAtomcenteredSymmetryFunctions2011}.
Such symmetry functions are rotationally, translationally, and permutationally invariant, and can capture both radial and angular features of the chemical environment within a cutoff radius $R_\textrm{c}$. 
Each chemical element has its own set of symmetry functions, and also its own neural network architecture and fitted parameters. The adoption of the atomic energy ansatz (Eq.~\ref{atomE}) in BPNNs makes the resulting ANN applicable to systems with an arbitrary number of atoms.

A GCNN~\cite{Duvenaud:2015ww, Kearnes:2016dt,schuttQuantumChemicalInsightsDeep2017} is a combination of a graph neural network and a convolutional neural network. 
A graph neural network considers the atoms as nodes and the pairwise interactions between them as weighted edges. 
Node and edge feature vectors are iteratively updated through e.g. a message passing function~\cite{gilmerNeuralMessagePassing2017}. The ingredient of convolution is often recognized as a learnable radial filter
which gathers information from neighboring atoms within a cutoff radius $R_\textrm{c}$ 
and creates a feature hierarchy. 
Because the generation of node features includes the element-specificity by construction 
(See Eq.~\ref{eq:pi} and Eq.~\ref{eq:ip_sum} in the next Section), 
GCNN has exactly the same subnet for each element.

\subsection{Pairwise interaction and interaction pooling}
\label{sec:pi_ip}
In PiNN, we describe both BPNN and GCNN with two abstractions (Fig.~\ref{fig:networks}): pairwise interaction operation (PI), and interaction pooling operation (IP).
We start by labeling atom $i$ with a set of scalar numbers, or
an atomic property $\vec{P}_i$. It is
sufficient to use the nuclear charge $Z_i$ or a one-hot embedding of the element as the starting point. 
Then, we create the pairwise interaction  $\vec{I}_{ij}$ as a function of the initial atomic properties of two atoms and their distance $r_{ij}$ (Eq.~\ref{eq:pi}).
\begin{equation}
  \label{eq:pi}
  \vec{I}_{ij}^t = \textrm{PI}\left(\vec{P}_i^t, \vec{P}_j^t, r_{ij}\right)
\end{equation}
where $t$ is an iterator. For BPNN, $t=0$ and for GCNN, $t+1$ goes up to the number of graph convolution (GC) blocks (Fig.~\ref{fig:networks}). 

The opposite of the PI operation is IP. Namely, this operation creates an atomic property from all the pairwise interactions associated with that atom. This is done
by passing the summation over all the pairwise interactions to another function called $\textrm{IP}$ (Eq.~\ref{eq:ip_sum}). The summation ensures the permutation invariance of the generated atomic property.
\begin{equation}
  \label{eq:ip_sum}
  \vec{P}_i^{t+1} = \textrm{IP}\left(\sum_j \vec{I}_{ij}^t\right)
\end{equation}

The combination of the PI and IP operations essentially creates an updated
atomic property, with information collected from neighboring
atoms. This is referred to as atomic fingerprints, continuous-filter
convolution or neural message passing in the
literature\cite{behlerConstructingHighdimensionalNeural2015,
  schuttSchNetDeepLearning2018,
  gilmerNeuralMessagePassing2017}. Through multiple GC blocks $\textrm{PI} + \textrm{IP} \rightarrow \textrm{PI} + \textrm{IP} \cdots$, $\vec{I}_{ij}$ also get updated in this process. The actual forms of the IP and PI functions are different in each ANN and this gives the freedom to create novel architectures (Fig.~\ref{fig:networks}).
 
\begin{figure}
  \centering
  \includegraphics[width=\columnwidth]{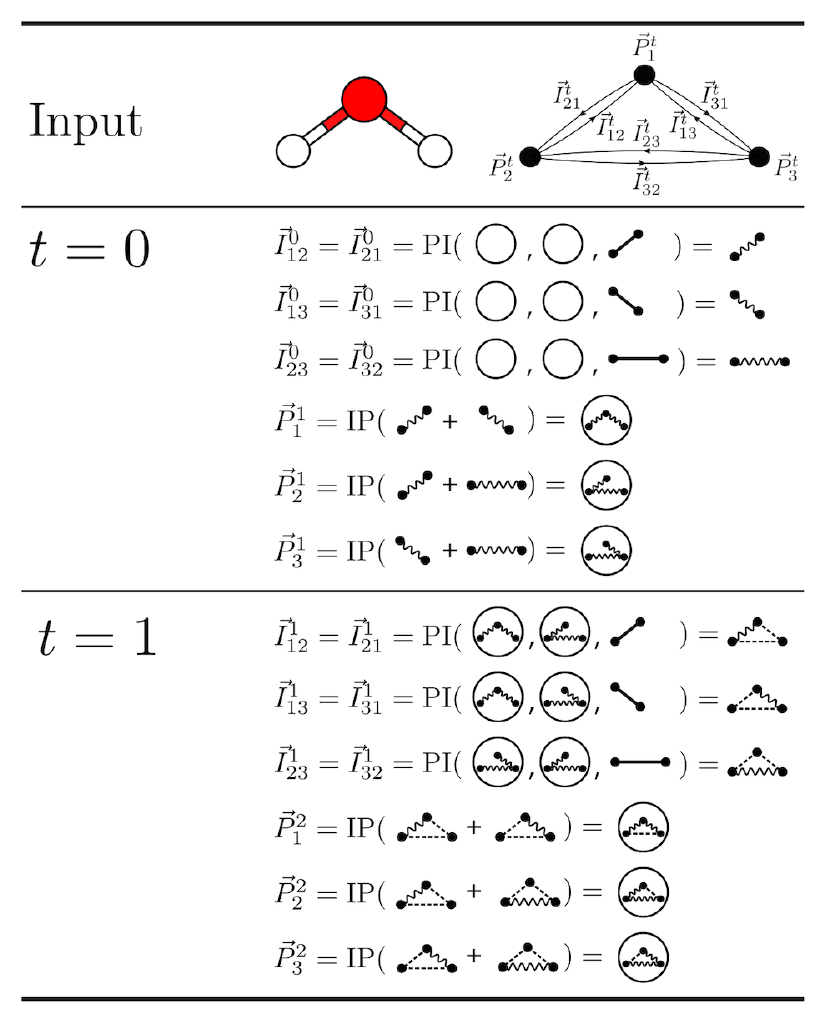}
  \caption[networks]{Illustrating how three-body representations for a water molecule are generated from two-body representations and graph convolution blocks in GCNN. $\vec{I}^t_{ij}$ and $\vec{P}^t_i$ are the pairwise interaction between atom $i$ and $j$ and the node feature vector of atom $i$ respectively. $t$ is the iterator in graph convolution blocks.}
  \label{fig:watex}
\end{figure}

To illustrate the many-body representation in GCNN, we use a single water molecule as an example to show how three-body interactions are generated from two-body representation, as shown in Fig.~\ref{fig:watex}. In the stage of embedding, the water molecule turns into a directed graph with node feature $\vec{P}^{t}_i$ and edge feature $\vec{I}^t_{ij}$ for $i=1,2,3$ and $j=1,2,3$. It is worth to note that $\vec{I}^t_{ij}$ is a high-dimensional feature vector and $\vec{I}^t_{ij}$ is not necessarily be the same as $\vec{I}^t_{ji}$.  At $t=0$, the PI operation will take node features $\vec{P}^0_i$, $\vec{P}^0_j$, and the bond $r_{ij}$ to generate the interaction $\vec{I}^1_{ij}$. Subsequently, the IP operation will update the node feature $\vec{P}^1_i$ by summing all interactions centered at atom $i$. $\vec{P}^1_i$ is equivalent to a vector of radial symmetry function values in BPNNs for atom $i$. At $t=1$, the PI operation generates a new interaction $\vec{I}^2_{ij}$ by repeating the same procedure, which is followed by another IP operation. What is important is to note that $\vec{P}^2_i$ is not only a three-body function but also a unique representation of the local chemical environment of atom $i$. Moreover, the pairwise interaction $\vec{I}^2_{ij}$ is already a three-body function in GCNN and the inclusion of explicit triplewise interactions $\vec{I}_{ijk}$ is possible but unnecessary.

To close this section, we note that although multiple GC blocks are capable of generating multi-body representations, there is no general proof regarding the uniqueness of GCNN representations. To the authors' knowledge, the closest example shows that GCNNs can be as powerful as the Weisfeiler-Lehman algorithm in detecting isomorphic graphs~\cite{Xu:2019ty}.

\subsection{Architecture of PiNet}
With the formalism of PI and IP, we designed and implemented a GCNN variant called PiNet. This network is motivated
by the aim to make the activations (pairwise interactions and atomic properties) of ANNs interpretable.

The main idea behind PiNet is to define the pairwise interaction as a function of distance, whose exact form depends on the atomic properties of both interacting atoms. In other words, the weight matrix $\mathbf{W}_{ij}$ used for generating the pairwise interaction $\vec{I}_{ij}$ depends on both atomic properties $\vec{P}_i$ and $\vec{P}_j$. This makes each component of the pairwise interaction $\vec{I}_{ij}$ to have different radial dependence, which is unique in PiNet and differs from the common approach of using a single radial-dependent
filter function (attention mask)
\cite{schuttSchNetDeepLearning2018, lubbersHierarchicalModelingMolecular2018, Unke:2019bp}.

In PiNet, the PI operation is split into three steps (Figure \ref{fig:networks}c): (i) expressing the interatomic distances in a radial basis $\vec{e}_{ij}$; (ii) activation through the PI ``layer'', which is a feed-forward NN generating a weight matrix $\mathbf{W}_{ij}$ from the atomic properties $\vec{P}_i$ and $\vec{P}_j$; (iii) activation through the II (Interaction to Interaction) layer, which is a feed-forward NN using the information from the previous two steps as input and generating the interaction $\vec{I}_{ij}$. 

The radial basis $\vec{e}_{ij}$ for
the pair of atoms $i$ and $j$ is calculated as
$n_\textrm{basis}$ Gaussian functions multiplied by a cutoff function $f_c$. 
\begin{equation}
\label{eq:gauss_basis}
\vec{e}_{ij} = f_c(r_{ij})\cdot [e^{-\eta(r_{ij}-r_{1})^2}, e^{-\eta(r_{ij}-r_{2})^2}, \cdots]
\end{equation}
The centers of the Gaussian functions $r_{1}, r_{2}, \cdots, r_{n_\textrm{basis}}$
are chosen to be evenly spaced  between 0 and the cutoff radius $R_\textrm{c}$, 
and the hyperparameter $\eta$ determines the width of the Gaussian functions.

We have chosen the cutoff function $f_\textrm{c}$ given in Ref.~\citenum{behlerAtomcenteredSymmetryFunctions2011}, which
ensures that the interaction and its gradient vanish at the cutoff radius $R_\textrm{c}$:
\begin{eqnarray}
f_{\textrm{c}}(r_{ij}) = \Bigg\{\begin{array}{cr}
                          0.5 \cdot [\cos(\frac{\pi r_{ij}}{R_c}+1)] & \text{for } r_{ij}<R_{\textrm{c}}\\
                          0 & \text{for } r_{ij} \ge R_{\textrm{c}}\\
                        \end{array}
\end{eqnarray}

The PI layer generates a weight matrix $\mathbf{W}_{ij}$ from the concatenated atomic properties $\vec{P}_i$ and $\vec{P}_j$:
\begin{equation}
\mathbf{W}_{ij} = \textrm{NN}^{\textrm{PI-layer}}([\vec{P}_{i}, \vec{P}_{j}]) \label{eq:weight}
\end{equation}

The II layer calculates $\vec{I}_{ij}$ by means of a different feed-forward NN
\begin{equation}
\vec{I}_{ij} = \textrm{NN}^\textrm{II-layer}(\mathbf{W}_{ij} \vec{e}_{ij}) \label{eq:PI}
\end{equation}
The radial basis $\vec{e}_{ij}$ decays to zero beyond the cutoff radius and all biases are set to zero in the II layer, which guarantees the smoothness of PES in PiNet in contrast to other approaches where the interaction is directly generated from the distance and the atomic properties~\cite{xieCrystalGraphConvolutional2018,chenGraphNetworksUniversal2019}.

It is worth mentioning that despite the fact that PiNet does not include a triplewise filter $\vec{I}_{ijk}$ in the current implementation, 
angular information is nonetheless captured through the iteration of the graph convolution block. 

After the PI operation, the updated atomic property $\vec{P}_i$ is calculated from all pairwise interactions $\vec{I}_{ij}$ as part of an IP operation (Eq. \ref{eq:ip_sum}). Before passing this atomic property on to the next GC block in the iteration, a PP (Property-to-Property) layer (another feed-forward NN) is used for the further refinement, see Figure~\ref{fig:networks}b.

\begin{figure*}[ht]
    \centering
    \includegraphics[width=0.8\textwidth]{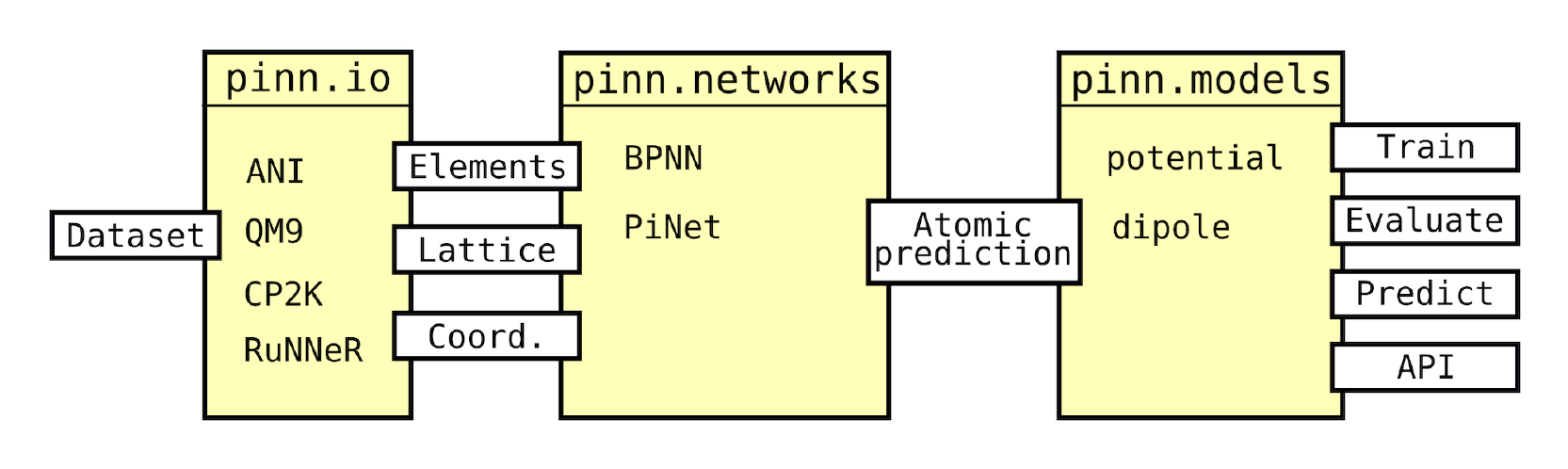}
    \caption{Illustration of the modularized structures in PiNN.}
    \label{fig:implement}
\end{figure*}

\section{Implementation and package features}
\label{sec:implementation}

\subsection{Modularized ANN}
PiNN is modularized so as to cater to the need of different uses.
The task of building an ANN is split into three stages:
dataset preparation, ANN definition, and model definition. 
As shown in Fig~\ref{fig:implement}, PiNN primarily consists of three modules:
input/output (io), networks, and models, which correspond to these three stages. The network does not specify the physical meaning of atomic predictions. 
That is instead defined through the loss function in a model. The model also contains hyper-parameters such as the learning rate and the regularization.  

This design enables a user to easily import an arbitrary dataset without touching the code base. Similarly, researchers interested in implementing a new ANN architecture could implement it as one of  ``networks'', and use the rest modules to test its performance. Finally, the ``models'' module could be extended to use the existing networks for different property predictions (e.g. dipole moments and partial charges in the Case Studies), or to interface with other external codes.

Notably, this modularized structure also decouples the implementation of common neural network building blocks, such as activation functions and optimization algorithms, from the ANN architecture. For example, a number of activation functions (e.g. tanh, logistic, or softplus) can be chosen in PiNN as adjustable hyperparameters of the network architecture.

\subsection{Dataset preparation}
\label{sec:dataset}
PiNN is implemented on top of TensorFlow's estimator
API\cite{abadi2016tensorflow}, which requires the
training data to be represented with the dataset class.

To enable easier utilization of data from different sources, we
provide the functionality of creating custom dataset
loaders. Given a list of data files and a reader function, the dataset
loader can be used to split the dataset, as is commonly
required for neural network training. Similar procedures can be applied
to trajectory files or databases. Further instructions for building
datasets are provided in the documentation\cite{pinn_documentation}.

Several types of dataset loaders are provided, such as for the QM9 dataset \cite{ramakrishnan2014quantum}, 
ANI-1 dataset \cite{smith_ani-1_2017},
Numpy\cite{numpy} formatted datasets, ASE\cite{HjorthLarsen:2017hn} databases, 
RuNNer-format \cite{behlerConstructingHighdimensionalNeural2015,behlerangewandte2017} datasets and CP2K\cite{Hutter:2013iea} trajectories in XYZ-format.

In addition, the dataset objects can be saved into TensorFlow's tfrecord file format,
for fast reading, and serving the dataset from a remote file
system. 

\subsection{Interfaces with ASE and AMS}
\begin{figure}
    \includegraphics[width=.45\textwidth]{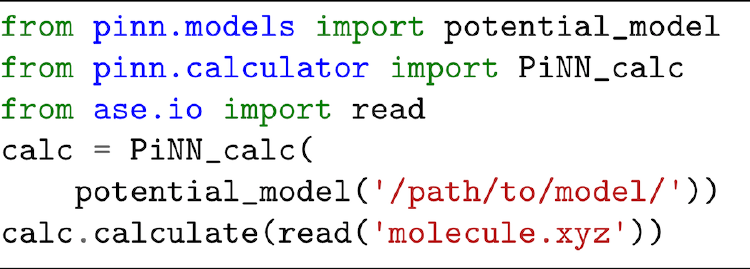}
    \caption{Code example for using a trained model as an ASE calculator.}
    \label{fig:code_block}
\end{figure}

PiNN comes with interfaces connecting the trained neural
network potential to ASE~\cite{HjorthLarsen:2017hn} through
its calculator class (see Fig~\ref{fig:code_block}), and to a development version of AMS\cite{AMS2019}
as an external engine.

Periodic boundary conditions are seamlessly supported. The neighbor list and pairwise distances of periodic system are efficiently calculated using a cell lists algorithm\cite{allen_computer_2017} that we implemented in TensorFlow. 

With either ASE or AMS, the PiNN implementations of BPNNs and PiNet can be used to run, for example, geometry optimizations and molecular dynamics (MD) simulations of both gas phase and condensed phase systems. The simulation trajectories can be visualized using their respective graphical user interfaces.

Moreover, the implementation of PiNN as an ASE calculator allows it easily to be interfaced with other codes. For example, the path integral molecular dynamics (PIMD) code i-PI\cite{Kapil:2019ju} can communicate with ASE calculators via a socket protocol, allowing PIMD simulations to be run with PiNN energies and forces.

\begin{figure*}[ht]
  \centering
  \includegraphics[width=0.95\textwidth]{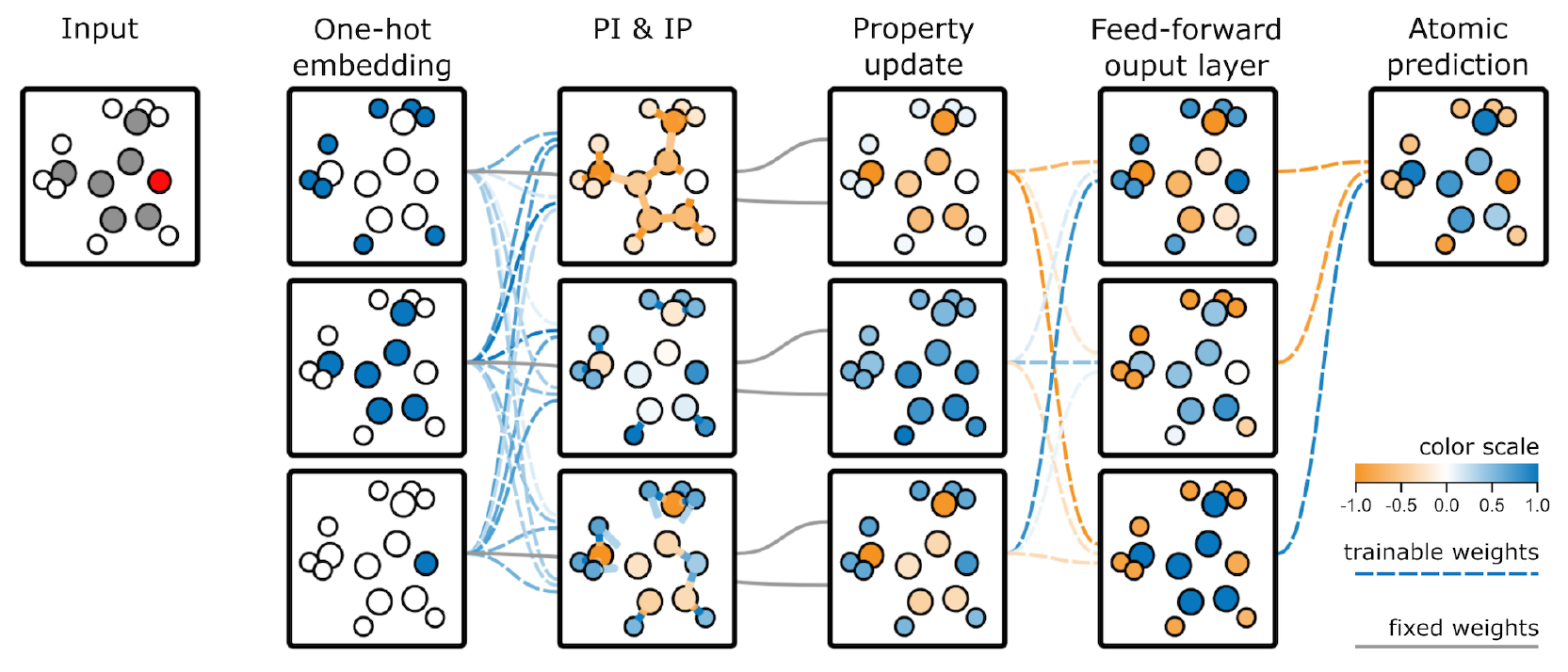}
  \caption{Visualization of a minimalistic PiNet for a 2,3-Dimethylfuran molecule generated directly with PiNNboard. For simplicity, only one graph convolution block without PP and II layers was used.
  The atoms and bonds in each box indicate the normalized activations of the atomic property
  and pairwise interaction respectively. 
  Dashed lines show the normalized trainable weights.
  Activation of pairwise interactions with an absolute value smaller than 0.3 is not shown for the sake of clarity.} 

  \label{fig:visualization}
\end{figure*}

\subsection{Pressure calculations} 
Computing the stress tensor requires attention, especially when the potential is not pairwise-additive
\cite{thompson2009general}. Although an ANN potential does not look like pairwise-additive, the stress tensor calculated using the pairwise form $\vec{F}_{ij} \cdot \vec{r}_{ij}$ yields the same result as the atomic form given in Ref.~\citenum{thompson2009general}. These results were also validated using the finite difference of the potential energy with respect to the change of the volume. Therefore, the instantaneous pressure $P(t)$ during an MD simulation is calculated as
\begin{equation}
    \label{pressure}
    P= \frac{1}{3\Omega}\left( \sum_i^N \frac{|\vec{p}_i|^2}{m_i} + \sum_{i}^N\sum_{j>i}^N \vec{F}_{ij} \cdot \vec{r}_{ij} \right)
\end{equation}
where
$\vec{F}_{ij} = \partial E_\textrm{tot}/\partial \vec{r}_{ij}$ is the derivative of the
potential energy with respect to the pairwise displacement vectors $\vec{r}_{ij}$~\cite{Martin:2013un},
$\vec{p}_i$ is the momentum of atom $i$, $m_i$ is the mass of that atom and $\Omega$ is the system volume. 
The first term on the right hand side of the equation is 
the ideal gas contribution and the second term is the virial pressure output from PiNN.

\subsection{Visualization of atomic neural networks with PiNNboard}
\label{sec:visualization}

It has been shown that a visualization of the activations can provide insights to their functions in convolutional neural network~\cite{zeiler_visualizingunderstandingconvolutional_2013}.
Such a visualization can also serve as a diagnostic tool to inspect and improve network architectures. Therefore, we developed a tool, PiNNboard, to visualize the activations of ANNs in atomic
or pair forms. PiNNboard was implemented as a plugin for Tensorboard -- TensorFlow's visualization toolkit~\cite{abadi2016tensorflow}.

Here we demonstrate PiNNboard with the minimalistic PiNet, trained on a subset of the 
QM9 dataset containing 50604 organic molecules consisting of only C, H and O atoms. The network was trained for 3 million steps on the internal energy at 0 K ($U_0$) and details about the training setups can be found in the Case Studies section below.

In Fig~\ref{fig:visualization}, the activations of this minimalistic PiNet for the testing molecule 2,3-Dimethylfuran are visualized using PiNNboard. Colored atoms indicates the contributions of each individual atom to atomic properties. Colored bonds between atoms indicates the pairwise interactions whose contributions are significant. Indeed, most of the interactions identified by the PI layer in PiNet can be recognized as covalent bonds in Fig~\ref{fig:visualization}. Interestingly, strong activations in the pairwise interaction are also observed between hydrogen atoms and their second neighbors in the case of the methyl groups. 

\begin{figure}[ht]
    \centering
    \includegraphics[width=\columnwidth]{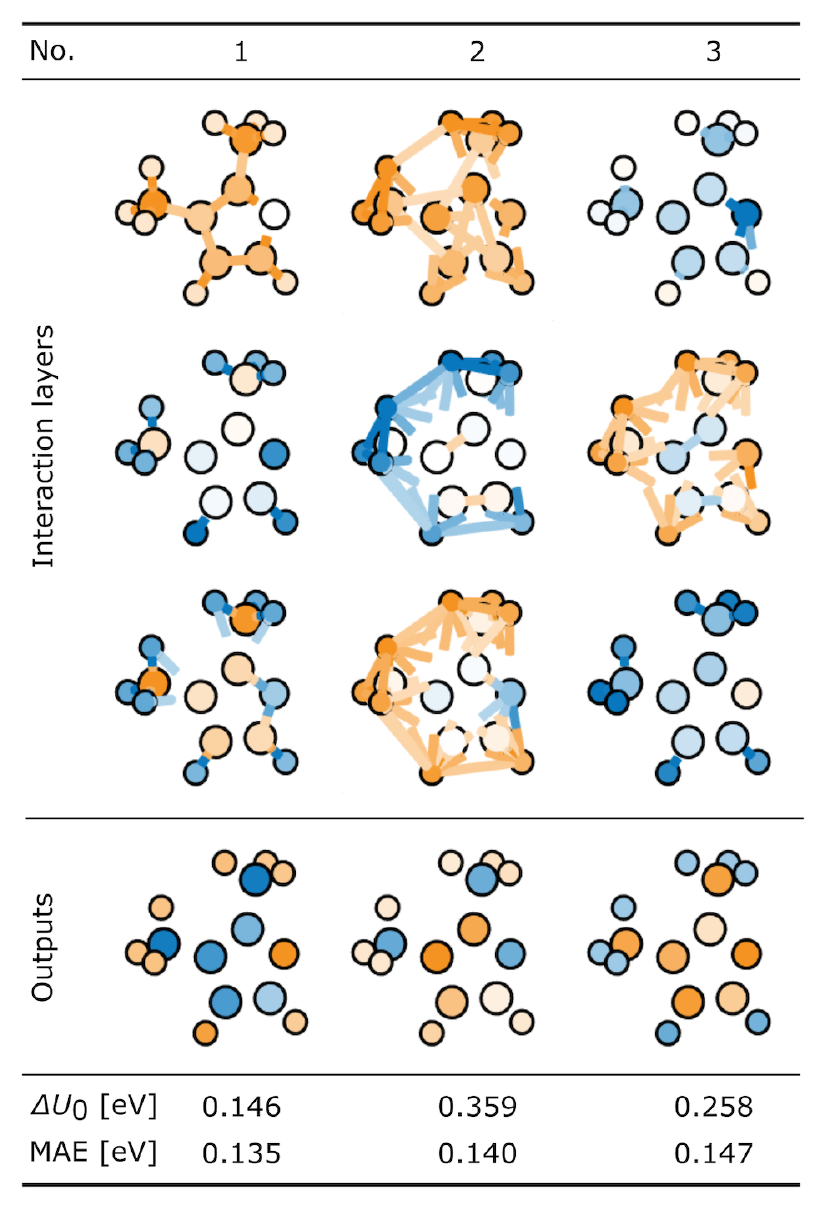}
     \caption{Activations of graph convolution and output layers after three independent trainings of the minimalistic PiNet shown in Fig~\ref{fig:visualization}. The prediction error in the $U_0$ of the testing molecule and the mean absolute error (MAE) on the validation set are also listed.
     }
    \label{fig:pinnboard_interactions}
\end{figure}

 Notably, three independently trained networks (Fig~\ref{fig:pinnboard_interactions}) reach quite different atomic energy partitions as well as different pairwise interactions for the testing molecules, in spite of their identical network structure, hyper-parameters and similar mean absolute error (MAE). Therefore, one needs to be careful when interpreting atomic energies extracted from ANNs, which has also been pointed out recently~\cite{gastegger2019}.
 
 We also notice that the trained network (No.1 in Fig~\ref{fig:pinnboard_interactions}) which has the smallest prediction error for the testing molecule also provides the best chemical interpretability. This suggests that not only is PiNet capable of providing a state-of-the-art performance but also the good outcome from PiNet can be rationalized in a chemically intuitive manner.

\section{Case studies}
\label{sec:case_study}
For all discussed benchmarks, a network with 5 graph convolution (GC) blocks is used. The parameters are given in Table~\ref{tab:params}. Hyperbolic tangent activation functions were used in all layers, except for the output layer where a linear activation function is used for output.

Unless otherwise stated, a 80:20 dataset splitting was used for case studies, which means 80\% of the structures of each dataset were randomly chosen to train the 
neural network, and the remaining 20\% were used for validation. The Adam \cite{kingma_adam:_2014}
optimizer in TensorFlow~\cite{abadi2016tensorflow} was used for gradient descent updates with a batch size of 100 samples,
the training rate was set to 0.0003 and decayed by a factor of 0.994 every 100000 steps,
other parameters were kept unchanged.
A gradient norm clipping strategy was employed to avoid exploding gradient problems
\cite{pascanu_difficulty_2012}.
The trainings were terminated after 1-3 million gradient descent steps, which typically takes a day with a single NVIDIA TITAN V GPU card.

\begin{table}[ht]
  \caption{Network parameters used in Case Studies.}
  \begin{threeparttable}
  \begin{tabularx}{\columnwidth}{ll|ll}
    \hline
    Layer & Architecture\tnote{a}  & Parameter & Value\\
    \hline
    PI  & [64]$\times$10 \tnote{b}  & $R_c$          & 4.5 $\textnormal{\AA}$\\
    II  & [64, 64, 64, 64] & GC blocks           & 5   \\
    PP  & [64, 64, 64, 64] & $n_\textrm{basis}$  & 10  \\
    Output & [64, 1]       & $\eta$              & 3.0 $\textnormal{\AA}^{-2}$ \\ 
    \hline                          
  \end{tabularx}
  \begin{tablenotes}
  \item[a] The layer architecture is denoted with the number of nodes in the
  hidden layers and in the output layer. [64, 64, 64, 64] means a neural network
  with three hidden layers, each with 64 nodes, and 64 output nodes.
  \item[b] The PI layer does not contain any hidden layer and the output dimension of the PI layer equals to the number of elements in $\mathbf{W}_{ij}$.
  
  \end{tablenotes}
  \end{threeparttable}
\label{tab:params} 
\end{table}

Because the purpose of PiNN is not to serve as a singular ANN architecture but to be used as a reliable and general-purpose library for further developments, we will not compare our results exhaustively with all other GCNN variants~\cite{schuttQuantumChemicalInsightsDeep2017, schuttSchNetDeepLearning2018, xieCrystalGraphConvolutional2018,lubbersHierarchicalModelingMolecular2018, Unke:2019bp, chenGraphNetworksUniversal2019,Zubatyuk:2019gp}. Instead, SchNet~\cite{schuttSchNetDeepLearning2018}, which pioneered the application of GCNN for modeling both molecules and materials will be quoted as the main reference to put our results into perspective.  

\subsection{QM9 dataset}
\label{sec:qm9}

The QM9 dataset
\cite{ramakrishnan2014quantum} 
is a dataset made up of 134k small organic molecules containing computed 
electronic, energetic and thermodynamic properties at B3LYP/6-31G(2df,p) level of theory
\cite{becke_densityfunctionalthermochemistry_1993, LEE:1988ub, stephens_initiocalculationvibrational_1994},
which is often used for benchmarking ANNs. As commonly done in the literature, 30054 structures 
which failed a consistency check were excluded during training and 
evaluation\cite{schuttSchNetDeepLearning2018,Unke:2019bp}. 

PiNet reaches a MAE of 0.012 eV for the prediction of internal energy at 0 K, in comparison with 0.014 eV from SchNet. 
It is worth to mention that the so-called chemical accuracy from thermo-chemistry measurements is about 0.043 eV.  

As an example of property predictions, we used PiNet to predict partial charges by regressing only the molecular dipole moment $\mu$: 
\begin{equation}
\label{eq:dip}
    \mu = \left| \sum_{i=1}^{N} \tilde{q_{i}} \vec{r}_{i} \right |
\end{equation}
where $\tilde{q_{i}}$ is the predicted partial charge on atom $i$. To ensure that the predicted total charge of each molecule is zero, we added a constraint term to the loss function.
 
By implementing this dipole model in PiNet,
we predicted the dipole moment with an MAE of 0.018 $D$ for the QM9 dataset. The network used to predict the dipole for the QM9 dataset was trained with a learning rate of 0.0001 and batch size of 200 structures instead. 

\begin{figure}[ht]
    \centering
    \includegraphics[width=\columnwidth]{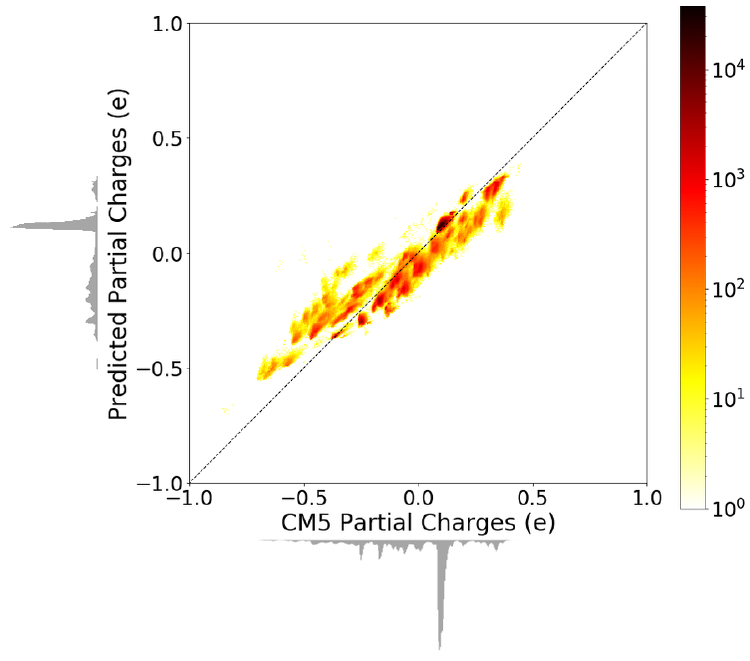}
     \caption{The correlation of the predicted partial charges from PiNet with those calculated from CM5 using the QM9 dataset.}
    \label{fig:QM9_charge_logarithmic}

\end{figure}

To further validate the PiNet dipole model, we also calculated partial charges
using the CM5 charge model, which has been parameterized to reproduce dipole moments from experiments or high-level quantum mechanical calculations~\cite{Marenich:2012et}. The CM5 charges were calculated at the B3LYP/6-31G(2df,p) level of theory using the Gaussian package\cite{gaussian}, matching the original conditions in which QM9 dataset was generated. 
Note that CM5 charges were not used in the training of PiNet. 
When comparing predicted partial charges from PiNet with those from CM5~\cite{Marenich:2012et},
a good correlation was found as shown in Fig.~\ref{fig:QM9_charge_logarithmic}, indicating that PiNet can generate physically meaningful partial charges.
This result is particularly encouraging in light of the fact that only the scalar dipole moment was used during the fitting~\cite{Sifain:2018fr}.

\subsection{Materials Project and Perovskite datasets}
\label{sec:mp}
We used the dataset (``MP-crystals-2018.6.1'') provided by MEGNet~\cite{chenGraphNetworksUniversal2019}
which contains DFT-computed energies and band gaps for 69640 crystals extracted from the Materials Project~\cite{Jain2013}.   Training of PiNet was done with 60000 crystal structures from this dataset. The trained PiNet leads to a MAE of 0.029 eV/atom for the prediction of formation energy on the test configurations, in comparison with that of 0.035 eV/atom from SchNet using the same number of structures in the training set. To put these numbers into perspective, we note that the accuracy of experimental measurement of formation energies is about 0.082 eV/atom~\cite{Kirklin:2015cr}. Thus, PiNet provides also a sub-chemical accuracy for materials modeling.

\begin{figure}[ht]
  \centering
  \includegraphics[width=\columnwidth]{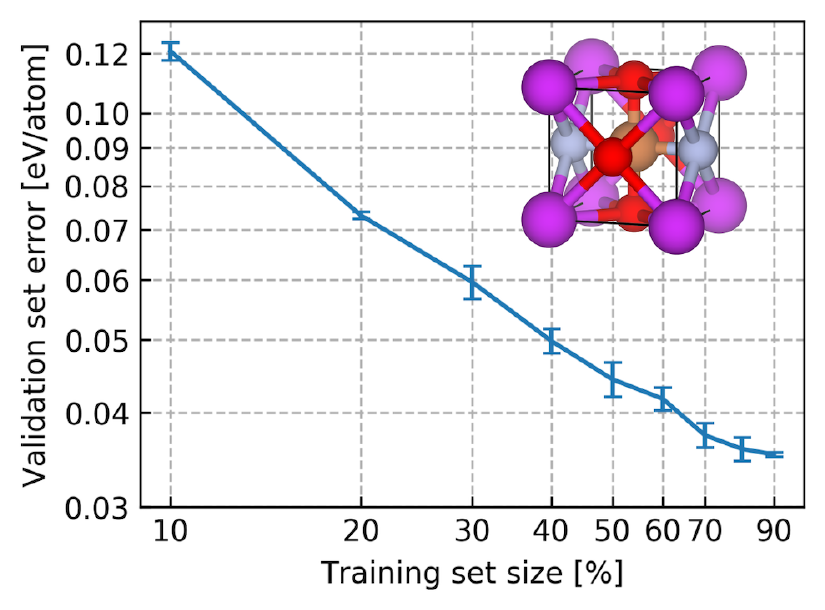}
  \caption{MAE of predicted energy above the hull from PiNet as a function of training configurations' size in the Perovskites dataset.}
  \label{fig:perovskite}
\end{figure}

In addition, PiNet was benchmarked on a dataset consisting of 18928 perovskite structures published by Castelli et al.~\cite{C1EE02717D}. 
The achieved MAE of the total energy respective to the convex hull (for the purpose of assessing the thermodynamics metastability) from the trained PiNet is 0.042 eV/atom with a 60:40 splitting of the dataset. This is in comparison with the same MAE obtained from Crystal Graph Convolutional Neural Network (CGCNN) with a 80:20 splitting instead~\cite{Xie:2018ga}. The learning curve of PiNet with this dataset is shown in Fig.~\ref{fig:perovskite} (in logarithm scale). 

\subsection{Liquid water dataset}
\label{sec:water}
Here we showcase the application of the BPNN implemented in PiNN to liquid water using the
dataset published by Morawietz et.al~\cite{Morawietz2019} based on the BLYP functional~\cite{becke88, LEE:1988ub}. To facilitate the training, we also augmented this dataset using the original BPNN implementation~\cite{Morawietz2019} (See Supporting Information for details). 
The set of symmetry functions were chosen to match the original ones~\cite{Morawietz8368}, however, hyperparameters such as the learning rate and optimizer are specific in PiNN.
MD simulations were performed with the Berendsen thermostat~\cite{Ber84} implemented in ASE and a
patched Berendsen barostat (to include the missing ideal gas contribution in Eq.~\ref{pressure}, for details see Ref.~\citenum{pinn_documentation}).

After the training, the BPNN implemented in PiNN reaches a root mean squared error (RMSE) of 7 meV/H$_2$O for energy and 60 meV/\AA~ for force components. These can be compared to 2 meV/H$_2$O and 70 meV/\AA~ for energy and force respectively, reported in Ref.~\citenum{Morawietz8368}. To validate this BPNN potential, we further carried out ab initio molecular dynamics (AIMD) simulations of the liquid water system at both NVT (constant particle, volume and temperature) and NPT (constant particle, pressure and temperature) ensembles with CP2K~\cite{Hutter:2013iea} and BLYP functional~\cite{becke88, LEE:1988ub}. Details of AIMD simulations can be found in the Supporting Information. 

\begin{figure}[ht]
  \centering
  \includegraphics[width=\columnwidth]{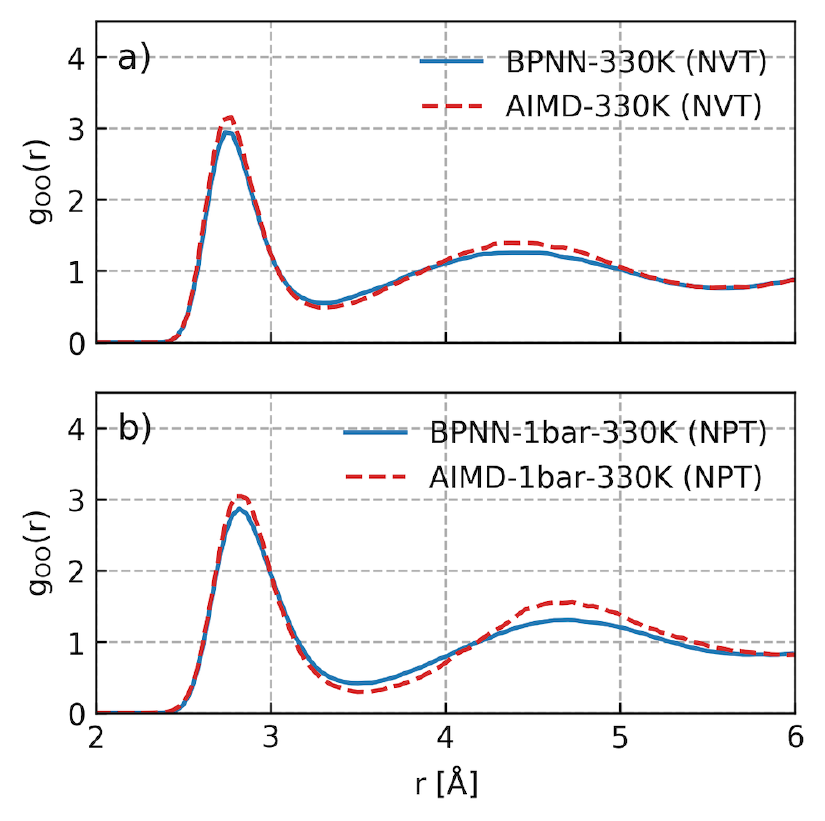}
  \caption{Calculated oxygen-oxygen radial distribution function of liquid water at a) 330 K (NVT) and b) 330 K and 1 bar (NPT) using BPNN implemented in PiNN and AIMD. The level of theory is BLYP.}
  \label{fig:goo_h2o}
\end{figure}

As shown in Fig~\ref{fig:goo_h2o}, the BPNN potential generated with PiNN reproduces well the structure of liquid water from AIMD simulations, particularly in the NVT ensemble. It is found that at 330 K and 1 bar, this BPNN potential predicts a density of 0.74(2) g/mL, in good agreement with that of 0.79(2) g/mL from AIMD simulations. Note that the AIMD simulations shown in Fig~\ref{fig:goo_h2o} were not used in the training and merely served the purpose of the cross-validation.

\subsection{Proton transfer reactions}
\label{sec:md}
Finally, we demonstrate that PiNet can be applied to
reactive MD simulations in which covalent bonds break and form.
In particular, we take the example of proton
transfer reactions in aqueous NaOH solutions.

We reused and slightly modified the dataset for NaOH(aq) solutions from
Ref.~\citenum{Hellstrom-JPCL-2016}, generated with the RPBE density functional\cite{RPBE} and Grimme's D3 dispersion correction \cite{Grimme-D3}.
This dataset was originally constructed
for use with BPNN and the resulting BPNN potential was tested for various thermodynamic and dynamical properties of NaOH solutions~\cite{Hellstrom-PCCP-2017,Hellstrom-JPCB-2018}. 

\begin{figure}[ht]
\centering
\includegraphics[width=0.45\textwidth]{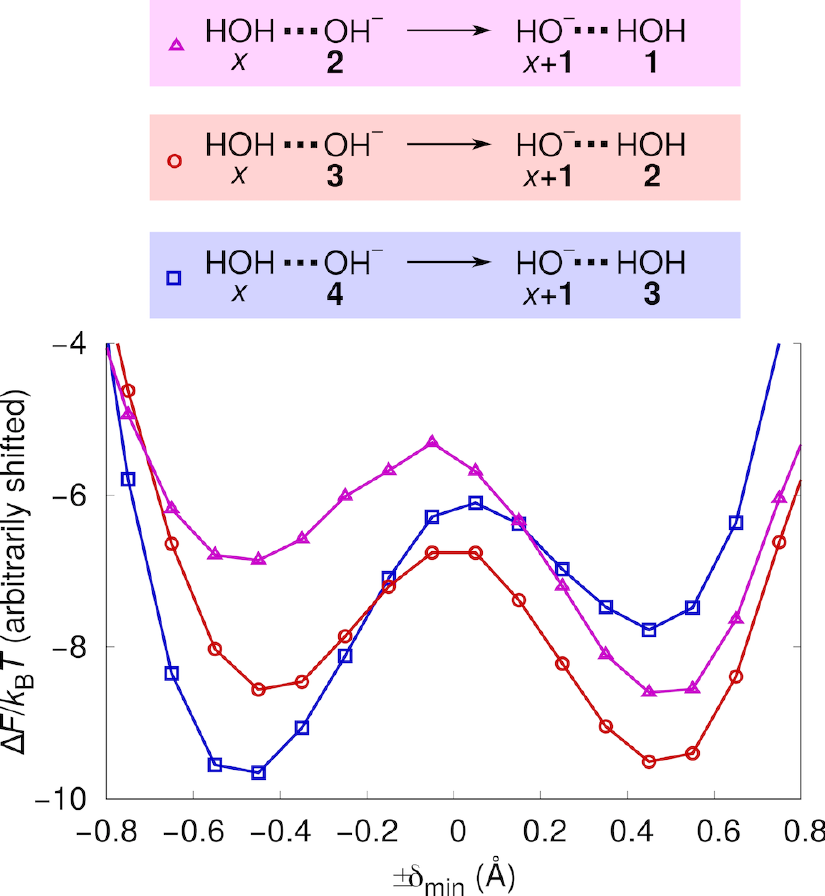}
\caption{\label{fig:presolvation} PiNet-calculated proton transfer free energy profile in 2.6 mol/L NaOH(aq) for \ce{OH-} and \ce{H2O} accepting different number of hydrogen bonds. The numbers beneath the molecules in the legend at the top refer to the total number of accepted hydrogen bonds. Here, we used a common definition of a hydrogen bond, for which the O-O distance is smaller than 3.5 \AA, and the hydrogen-bonding angle is smaller than 30 degrees. 
}
\end{figure}

In this case, we used smaller layers in PiNet (with 16 nodes per layer rather than 64) as compared to the other case studies to prevent overfitting, and optimized the parameters primarily to minimize the error in the predicted forces. We obtained an excellent RMSE of 0.11 eV/\AA\ for the force components for both the training and validation sets, indicating that the fit did not suffer from overfitting.

Environment-dependent proton transfer
free energy profiles for NaOH solutions were calculated with PiNet (Figure~\ref{fig:presolvation}).  For
each \ce{OH-}, the proton transfer coordinate $\delta_\textrm{min}$ is
calculated as the difference in length between the hydrogen bond along which
the proton is transferred, and the covalent O-H distance for the bond that
becomes broken \cite{Tuckerman-Nature-2002}. Corresponding equilibrium MD simulations were run using an interface with the Amsterdam Modeling Suite (AMS)~\cite{AMS2019} and technical settings of MD simulations were chosen to be the same as in Ref.~\citenum{Hellstrom-JPCL-2016} for the sake of comparison.

Figure~\ref{fig:presolvation} illustrates how the free-energy landscape for proton
transfer depends on the local hydrogen-bonding environments around \ce{OH-} and
\ce{H2O}.  Just as revealed in the previous work using BPNNs \cite{Hellstrom-JPCL-2016} and ab initio simulations \cite{Tuckerman-Nature-2002}, proton transfer occurs predominately via a presolvation mechanism: \ce{OH-} mostly accepts four hydrogen bonds in its equilibrium structure, but the forward PT barrier is quite high in this case (blue curve). Instead, if via a hydrogen bond fluctuation the \ce{OH-} only accepts three hydrogen bonds (red curve), then the forward PT barrier is much smaller. 

\begin{figure}[ht]
\centering
\includegraphics[width=0.45\textwidth]{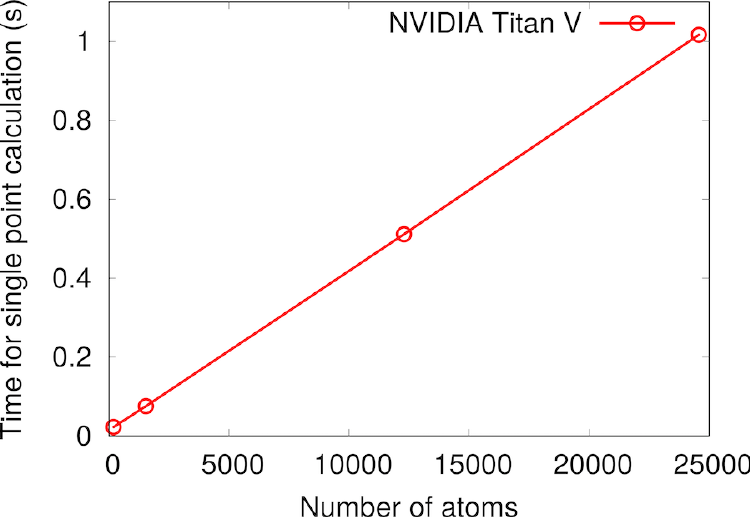}
\caption{\label{fig:scaling}Computational cost as a function of system size, for the NaOH PiNet model with a single GPU card.
}
\end{figure}

Figure~\ref{fig:scaling} shows the time required to evaluate the energy and forces for different system sizes of liquid water (averaged over 100 samples), using the PiNet model parameterized for NaOH. This highlights one of the appealing features of ANNs in which the computational cost grows linearly with the number of atoms.

\section{Conclusion}
\label{sec:conclusion}
Here we present PiNN -- an open-source Python library for building ANNs of molecules and materials. In the current version of PiNN, BPNN and our GCNN variant PiNet have been implemented and benchmarked against several publicly available datasets as well as in-house data. 

Built with the TensorFlow framework, PiNN allows for fast
training of ANNs with GPUs. PiNN interfaces with ASE and AMS and provides several useful package features such as analytical stress tensor calculations and a visualizer of trained ANNs called PiNNBoard.

With modularized building blocks, PiNN can be used not only as a standalone package but also as a chain of tools to develop novel ANNs. In this work, we showed how such ANNs can be used for approximating potential energy surfaces, allowing for fast and accurate reactive MD simulations, or for directly predicting several different physico-chemical properties of molecules and materials, such as dipole moments and partial charges. 

In the current implementation of PiNN, the primary focus is on predicting atomic properties. In the near future, we aim to include predictions of electronic properties such as polarization as well as the coupling to the external field in periodic systems. These extensions will be quite important for modeling electrochemical systems with finite field MD simulations~\cite{Zhang:2019ce, Zhang:2016ho}. Moreover, we are interested in incorporating active learning and on-the-fly potential generation.

Last but not least, PiNN will take advantage of the evolving TensorFlow ecosystem which allows to access novel optimization methods implemented in its peripheral packages such as K-FAC (Kronecker-Factored Approximated Curvature)~\cite{Martens:2015wb} for fast training of ANNs.

\begin{acknowledgement}
CZ is grateful to Uppsala University for a start-up grant, to the Swedish Research Council for a starting grant (No. 2019-05012) and to the Swedish National Strategic e-Science program eSSENCE for funding. MH received funding from the European Union's Horizon 2020 research and innovation programme under grant agreement No 798129. Supports from NVIDIA Corporation GPU grant program and Google Cloud Platform (GCP) research credits award are also gratefully acknowledged. Part of the simulations were performed on the
resources provided by the Swedish National Infrastructure for Computing (SNIC) at UPPMAX and PDC. 
\end{acknowledgement}

\begin{suppinfo}

Details of liquid water dataset augmentation  and AIMD simulations of liquid water.

\end{suppinfo}


\providecommand{\latin}[1]{#1}
\makeatletter
\providecommand{\doi}
  {\begingroup\let\do\@makeother\dospecials
  \catcode`\{=1 \catcode`\}=2 \doi@aux}
\providecommand{\doi@aux}[1]{\endgroup\texttt{#1}}
\makeatother
\providecommand*\mcitethebibliography{\thebibliography}
\csname @ifundefined\endcsname{endmcitethebibliography}
  {\let\endmcitethebibliography\endthebibliography}{}

\end{document}


\newpage

\section{Liquid water dataset augmentation}
Morawietz et al.~\cite{Morawietz8368} parameterized several Behler-Parrinello neural networks (BPNNs) with datasets containing energies and atomic forces, calculated at several different levels of theory, for structures of ice and liquid water. In the present work, we reused their
dataset~\cite{Morawietz2019} calculated at the BLYP level of theory to
validate the BPNN implementation in PiNN. To aid the fitting
procedure, we augmented the dataset with 2841 structures of liquid
water distributed in the same density range. We used the original BPNN from Morawietz et al.\cite{Morawietz8368} implemented in the RuNNer code\cite{behlerConstructingHighdimensionalNeural2015,behlerangewandte2017} to run molecular dynamics simulations of liquid water, and extracted random snapshots from these simulations for inclusion in the training set. The reference data, to which we trained our own BPNN implemented in PiNN, thus contained energies and forces calculated using DFT as well as using the original BPNN from Morawietz et al., which has been shown to have first-principles quality. 

\section{AIMD simulations of liquid water}

The electronic structure of liquid water was solved applying DFT in the BLYP approximation~\cite{becke88, LEE:1988ub} as implemented in CP2K~\cite{Hutter:2013iea}. Triple-$\zeta$ basis sets with two additional polarization functions (TZV2P) and a charge density cutoff of 600 Ry were used. Core electrons were taken into account using the dual-space Goedecker-Teter-Hutter (GTH) pseudopotentials\cite{Goedecker:1996ve}. The model system consisted  of 64 water molecules in a cubic box of length  12.432~\AA~as the initial condition.

For the NVT simulation, Bussi-Donadio-Parrinello thermostat~\cite{Bussi:2008wu} was applied to keep the temperature at 330K. For the NPT simulation, Martyna -Tuckerman-Tobias-Klein barostat~\cite{Martyna:1996ga} was employed as well. In both simulations, the time-step was chosen to be 0.5 fs. Trajectories were collected for 20 ps in each case, which has been shown to be sufficient to obtain a reliable estimation of the density~\cite{Schmidt:2009tf}.

\providecommand{\latin}[1]{#1}
\makeatletter
\providecommand{\doi}
  {\begingroup\let\do\@makeother\dospecials
  \catcode`\{=1 \catcode`\}=2 \doi@aux}
\providecommand{\doi@aux}[1]{\endgroup\texttt{#1}}
\makeatother
\providecommand*\mcitethebibliography{\thebibliography}
\csname @ifundefined\endcsname{endmcitethebibliography}
  {\let\endmcitethebibliography\endthebibliography}{}